\documentstyle[aps,prb,multicol,psfig]{revtex}
\begin{document}

\title{Quasiclassical approach to the weak levitation of extended states in
the quantum Hall effect}

\author{M. M. Fogler}

\address{School of Natural Sciences, Institute for Advanced Study,
Olden Lane, Princeton, NJ 08540}

\date{
\today\
}

\maketitle

\raisebox{90pt}[0pt][0pt]{\mbox{\hspace{5in}Preprint IASSNS-HEP-97/132}}

\vspace{-0.3in}

\begin{abstract}

The two-dimensional motion of a charged particle in a random potential
and a transverse magnetic field is believed to be delocalized only at
discrete energies $E_N$. In strong fields there is a small positive
deviation of $E_N$ from the center of the $N$th Landau level, which is
referred to as the ``weak levitation'' of the extended state. I
calculate the size of the weak levitation effect for the case of a
smooth random potential re-deriving earlier results of Haldane and Yang
[\prl {\bf 78}, 298 (1997)] and extending their approach to lower
magnetic fields. I find that as the magnetic field decreases, this effect
remains weak down to the lowest field $B_{\rm min}$ where such a
quasiclassical approach is still justified. Moreover, in the immediate
vicinity of $B_{\rm min}$ the weak levitation becomes additionally
suppressed. This indicates that the ``strong levitation'' expected at
yet even lower magnetic fields must be of a completely different origin.

\end{abstract}

\begin{multicols}{2}

The levitation of the extended states in the quantum Hall effect has
been proposed by Khmelnitskii~\cite{Khmelnitskii} and
Laughlin~\cite{Laughlin} (KL) almost fifteen years ago. Based on scaling
arguments (see Ref.~\onlinecite{Huckestein} for review) KL
further suggested that the energies $E_N$ of the extended states can be
obtained by solving the equation
\begin{equation}
\sigma_{xy}(E_F = E_N) = (N + 1/2)\,\frac{e^2}{2\pi \hbar},
\label{scaling}
\end{equation}
where $\sigma_{xy}$ is the unrenormalized (short length scale)
Hall conductance, $E_F$ is the Fermi energy, and $N$ runs through
the set of integer values $N = 0, 1, \ldots$. Using the Drude-Lorentz
formula
\begin{equation}
      \sigma_{xy} = \frac{e^2}{m}\,
      \frac{\omega_c}{\omega_c^2 + \tau^{-2}}\, n(E_F),
\label{Drude}
\end{equation}
for the left-hand side, they obtained~\cite{Khmelnitskii}
\begin{equation}
   E_N = (N + 1/2)\, \hbar \omega_c\, [1 + (\omega_c \tau)^{-2}],
\label{KL_full}
\end{equation}
where $\omega_c$ is the cyclotron frequency, $\tau$ is
the zero-field momentum relaxation time, and $n$ is the electron density.
Thus, in strong fields where $\omega_c\tau \gg 1$ the energy $E_N$ of
the $N$th extended state is close to the center of the $N$th Landau
level $E_N^\infty = \hbar\omega_c(N + 1/2)$. As the magnetic field
decreases, $E_N$ floats upward with respect to $E_N^\infty$, so
that the relative deviation $\delta E_N/E_N^\infty$ increases
\begin{equation}
 {\delta E_N}/{E_N^\infty} \equiv
 ({E_N - E_N^\infty})/{E_N^\infty} = (\omega_c \tau)^{-2},
\label{KL}
\end{equation}
I will call the regimes $\delta E_N/E_N^\infty \ll 1$ and $\delta E_N/E_N^\infty
\gg 1$ a ``weak'' and a ``strong'' levitation regimes, respectively. The
regime of strong levitation is, of course, the most interesting.
Unfortunately, this regime is also the hardest one to study. Up to date
there is no progress in analytical treatment of this problem; as for the
original arguments,~\cite{Khmelnitskii} they suffer from the absence of
completely satisfactory derivation of the scaling
laws.~\cite{Huckestein}

Not long ago, the study of the weak levitation phenomenon has been
pioneered by Shahbazyan and Raikh~\cite{Shahbazyan_95} and then
continued by several other
groups.~\cite{Kagalovsky,Liu,Gramada,Haldane_97,Sheng} The advances in
the analytical treatment of the problem are due to the existence of a
transparent physical picture of localization in sufficiently strong
magnetic field and a smooth random potential.~\cite{Drift_2d} I will
start with briefly recalling this picture, which leads to an approximate
equality $E_N \approx E_N^\infty$. Then I will review the arguments of
Haldane and Yang~\cite{Haldane_97} (HY) who identified the leading
contribution to $\delta E_N = E_N - E_N^\infty$. Finally, I will
present my own results.

The first simplification of the strong magnetic field limit comes from
the approximate separation of the electron's motion into a fast rotation
along the cyclotron orbit and a much slower dynamics of its guiding
center. Unfortunately, the guiding center coordinates (unlike the
coordinates of the particle itself) can not have definite values
simultaneously. The characteristic uncertainty in the guiding center
position is of the order of the magnetic length $l = \sqrt{\hbar /
m\omega_c}$. And here the smoothness of the random potential brings the
second crucial simplification: as long as the correlation length $d$ of
the potential is much larger than $l$, this uncertainty can be ignored.
As a result, there are two approximate integrals of motion: the energy
of the cyclotron motion, i.e., the kinetic energy, and the energy of the
guiding center degree of freedom, which is essentially the potential
energy. In this approximation the guiding center is permanently bound to
a certain level line $U(x, y) = {\rm const}$ of the random potential
$U(x, y)$ and performs a slow drift along such a line. The extended
states correspond to {\em unbounded\/} level lines.~\cite{Drift_2d} In
fact, for a wide class of potentials statistically symmetric under the
sign change, the unbounded level line (percolation contour) is unique
and is at zero energy.~\cite{Isichenko} Since the total energy of the
electron is equal to the sum of the energies of the cyclotron and the
drift degrees of freedom, and the cyclotron energy is quantized in
$\hbar\omega_c$ quanta, in this approximation $E_N$ is equal to
$E_N^\infty$.

The calculation of the the weak levitation correction requires taking
into account the Landau level mixing. HY demonstrated that
such a mixing simply modifies the form of the potential in which the
guiding center drifts. The new potential is not statistically
symmetric under sign change. In fact, its percolation level is at higher
energy.~\cite{Haldane_97} Correspondingly, $E_N$ is larger than $E_N^\infty$.

Using the quantum-mechanical perturbation theory, HY found that
\begin{equation}
 \delta E_N \sim (N + 1/2) (W^2 / \hbar\omega_c) (l / d)^4,
\label{HY}
\end{equation}
where $W$ is the rms amplitude of the random potential. I retained only
the first term in the perturbation series obtained by HY. The next
term is smaller if $d \gg l N^{1/2}$. In addition, HY require that
$\hbar\omega_c \gg W$. Denote by $R_c$ the classical cyclotron radius at
energy $E_N$. It is easy to see that $R_c = \sqrt{2 N + 1}\, l$, so that
Eq.~(\ref{HY}) becomes
\begin{equation}
 \delta E_N / E_N^\infty \sim (W / E_N)^2 (R_c / d)^4,
\label{Haldane}
\end{equation}
and the condition $d \gg l N^{1/2}$ is simply $R_c / d \ll 1$. It
immediately hits the eye that HY's result is expressed in terms of
purely classical quantities, given the particle's energy is equal to $E_N$.
Note also that the size of the effect is different from KL's
formula~(\ref{KL}). For instance, it is
much larger provided that $R_c/ d \gg W / E_N$. For a weak random
potential, $W \ll E_F \simeq E_N$, this inequality can be met simultaneously
with $R_c / d \ll 1$.

\begin{figure}
\centerline{
\psfig{file=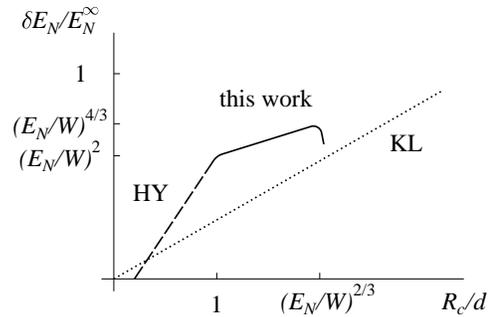,width=2.5in,bbllx=140pt,bblly=280pt,bburx=470pt,bbury=505pt}
}
\vspace{0.1in}
\setlength{\columnwidth}{3.2in}
\centerline{\caption{
The sketch of the dependence of the relative size of
the weak levitation effect $\delta E_N / E_N^\infty$ at
the topmost Landau level on the ratio $R_c/d$ (see text).
Labels ``HY,'' ``this work,'' and ``KL'' refer to
Eqs.~(\protect\ref{Haldane}), (\protect\ref{MF}), and
(\protect\ref{KL}), respectively. Both axes are in log-scale.
\label{plot}
}}
\end{figure}

Naively, one might think that the weak levitation mechanism of HY stops
functioning in lower magnetic fields where $R_c \gg d$. This turns out
not to be true; however, the dependence of the weak levitation effect on
magnetic field becomes slower,
\begin{equation}
 \delta E_N / E_N^\infty \sim (W / E_N)^2 (R_c / d) \sim
 (\omega_c \tau)^{-1}.
\label{MF}
\end{equation}
[The final expression follows from $\tau \sim (d / v_F) (E_F / W)^2$].
Equation~(\ref{MF}) is the central result of this paper. It is
represented graphically in Fig.~\ref{plot} together with the previous
two. The plot should be understood as the dependence of the quantity
$\delta E_N / E_N^\infty$ at the topmost Landau level on the ratio $R_c
/ d$. In other words, I assumed that the Fermi energy $E_F$ is fixed but
the magnetic field is changing. For each value of the magnetic field one
has to choose $E_N$ closest to $E_F$. Of course, discreteness of $N$
leads to some fine details on the curve in Fig.~\ref{plot}. Such details
are insignificant for large $N$, which is assumed to be the case for the
most points on the plot.

As one can see from Eq.~(\ref{MF}), $\delta E_N / E_N^\infty$ monotonically
grows as the magnetic field decreases. Naturally, one would like to know
if it ever becomes of the order of one. The answer is negative: even at the
lowest magnetic field $B_{\rm min}$ where the present approach is
justified, the quantity $\delta E_N / E_N^\infty$ is still small. To
verify that one needs to know what $B_{\rm min}$ is. Clearly,
$B_{\min}$ is the largest of the two fields, at which the two
simplifying considerations mentioned in the beginning of the paper breaks
down. One is the field $B_c$ where the separation into the cyclotron and
drift motion ceases to be valid, and the other is the field where
the quantum uncertainty ($\sim l$) of the guiding center position becomes
of the order of $d$. The crossover field $B_c$ was calculated in
Ref.~\onlinecite{Fogler}. It corresponds to the point
$R_c/d \sim (E_F / W)^{2/3}$ where the characteristic frequency of
the drift motion becomes of the order of $\omega_c$. Combining this with
the other condition, one obtains the largest value of the ratio $R_c / d$
where the calculation is still valid,
\begin{equation}
 (R_c / d)_{\rm max} = \min\,\{(E_F/W)^{2/3},\, k_F d\}
\label{R_c_d_max}
\end{equation}
($k_F$ is the Fermi wavevector in zero field).
Substituting this value into Eq.~(\ref{MF}) and keeping in mind that
$W \ll E_F$, one obtains that $(\delta E_N / E_N^\infty)_{\rm max} \ll 1$
(see also Fig.~\ref{plot}). 
%
%

Let us now turn to the derivation of Eqs.~(\ref{Haldane}) and
(\ref{MF}). I am going to show that the effect is completely classical and
therefore the constraint $W \ll \hbar\omega_c$ imposed by HY is
extraneous. The relevant condition is just $R_c/d \ll (R_c/d)_{\rm
max}$. Also, in contrast to the
quantum-mechanical treatment of Ref.~\onlinecite{Haldane_97}, my
instrument will be the {\em classical\/} perturbation theory. For this
reason I will drop the subscript ``$F$'' in $E_F$. To ensure continuity
with the previous paper,~\cite{Fogler} I will assume that the magnetic field
is in the negative $\hat{z}$-direction, so that the cyclotron
gyration is clockwise and the guiding center coordinates are given by
%
\[
\rho_x = x + (v_y / \omega_c),\:\:\:
\rho_y = y - (v_x / \omega_c),
\]
%
where $x$ and $y$ are the coordinate of the electron, and $\bbox{v} =
-v\,(\sin\theta, \cos\theta)$ is its velocity.
The most convenient form of the
equation of motion for $\bbox{\rho}$ is obtained if $U$ is re-expressed
in terms $\bbox{\rho}$ and $\theta$ only, i.e., as a new function
%
\[
V(\bbox{\rho}, \theta) = U[\rho_x + (v / \omega_c) \cos\theta,
\rho_y - (v / \omega_c) \sin\theta],
\]
%
where $v =\sqrt{2 (E - V) / m}$ because of the energy conservation.
The equation of motion for $\bbox{\rho}$ is
\begin{equation}
d\bbox{\rho} / d\theta = \frac{1}{m \omega_c^2}
[\hat{z} \times \bbox{\nabla}_{\rho} V(\bbox{\rho}, \theta)].
\label{eom whole}
\end{equation}
This equation is of the Hamiltonian form with $\theta/\omega_c$,
$\rho_y$, $m\omega_c\rho_x$, and $V$ being the time
variable, the canonical coordinate, momentum, and the Hamiltonian
function, respectively. The return to the original time variable can be
accomplished by means of the equation
%
\[
\dot\theta = \omega_c + \hat{z}\,
[\bbox{v} \times \bbox{\nabla}U] / (m v^2).
\]
%
It is convenient to expand $V(\bbox{\rho}, \theta)$ in Fourier series,
\begin{equation}
V(\bbox{\rho}, \theta) = \sum_{k = -\infty}^\infty
V_k(\bbox{\rho})\,e^{-i k \theta}.
\label{Fourier}
\end{equation}
If $|k| \lesssim R_c / d$, the absolute value of $V_k$ is
of the order of $W_0 = W (d / R_c)^{1/2}$; otherwise,
it is much smaller. Substituting Eq.~(\ref{Fourier}) into
Eq.~(\ref{eom whole}), one obtains
\begin{equation}
d\bbox{\rho} / d\theta = \frac{1}{m \omega_c^2}
\sum_{k = -\infty}^\infty\,
[\hat{z} \times \bbox{\nabla}_{\rho} V_k(\bbox{\rho})]\, e^{-i k \theta}.
\label{eom}
\end{equation}
If one retains only the $k = 0$ term, then the right-hand side does will
not depend on $\theta$ and thus the guiding center motion will decouple
from the cyclotron one. In this approximation the guiding center
performs the drift along the level lines of the potential $V_0$. In the
$R_c \ll d$ limit $V_0$ is very close to the original potential $U$ in
agreement with the qualitative picture given above.
If $R_c$ is larger than $d$, then quite a few $k\neq 0$ terms
are of the same magnitude as $k = 0$ one.
In this case one can not simply ignore them; however, they can be made
smaller by means of series of canonical (or almost canonical)
transformations.~\cite{Averaging}
Each consecutive transformation reduces the oscillating terms by a factor
of the order of $\gamma \equiv W_0 / (m \omega_c^2 d^2) \ll 1$. In the
end they become suppressed by a factor $\exp(-{\rm const} / \gamma)$. 
This program can be realized only if $\gamma \ll 1$. Equation
$\gamma = 1$ thus determines the magnetic field $B_c$ (see above) where the
crossover from the adiabatic drift to the random walk of the guiding
center occurs.~\cite{Fogler}

For calculation of $\delta E_N$ to the first nonvanishing order in
$\gamma$ only one such a transformation suffices. Let $p$ and $q$ be the
new canonical coordinates after the transformation. Define
the ``renormalized'' guiding center coordinates, $\rho_x^{(1)} = p /
(m\omega_c)$ and $\rho_y^{(1)} = q$. It is easy to see that
$\bbox{\rho}^{(1)}$ has the following form
\begin{equation}
\bbox{\rho}^{(1)} = \bbox{\rho} + \frac{1}{m\omega_c^2}
\sum_{k\neq 0} \frac{1}{i k}\,
[\hat{z} \times \bbox{\nabla}_{\rho} V_k] + O(\gamma^2 d).
\label{rho_1}
\end{equation}
The $\theta$-independent term $V_0$ in the
Hamiltonian function is transformed into $V^{\rm eff}$ given by
%
\[
V^{\rm eff} = V_0 + \frac{1}{m\omega_c^2}
\sum_{k = 1}^\infty \frac{\hat{z}}{i k}\,
[\bbox{\nabla} V_{-k} \times \bbox{\nabla} V_{k}] + O(\gamma^2 W_0).
\]
%
Define
%
\[
U_k(\bbox{\rho}, K) \equiv \oint\frac{d \phi}{2 \pi}\,e^{-i k \phi}
  U[\rho_x + R \cos \phi, \rho_y + R \sin \phi],
\]
%
where $K$ has the meaning of the kinetic energy and $R = \sqrt{2 K / m
\omega_c^2}$ of the corresponding cyclotron radius. Note a useful formula
\begin{equation}
\tilde{U}_k(\bbox{q}, K) = i^k e^{-i k \theta_k} J_k(q R)
  \tilde{U}(\bbox{q}),
\label{U_k} 
\end{equation}
where tilde symbolizes the Fourier transform, $\bbox{q} = q
(\cos\theta_k, \sin\theta_k)$, and $J_k$ is the Bessel function.

It is easy to see that
\begin{eqnarray}
&&V^{\rm eff}(\bbox{\rho}, K) = U_0 + 
\frac{1}{m \omega_c^2} \sum_{k = 1}^\infty (Y_k - Z_k) + O(\gamma^2 W_0),
\label{V_eff}\\
&&Y_k = \displaystyle \frac{\hat{z}}{i k}\,
[\bbox{\nabla} U_{-k} \times \bbox{\nabla} U_{k}],\:\:\:\:
Z_k =  \displaystyle \frac{1}{R} \frac{\partial}{\partial R}
|U_k|^2.
\label{Z_k}
\end{eqnarray}
%
The obtained expression agrees with the effective potential of HY
in the limit $R_c \ll d$. To see that one has to
quantize the kinetic energy $K = E - V^{\rm eff} =\hbar\omega_c (N +
1/2)$ and keep only $k = 1$ and $k = 2$ terms, which dominate the sum in
this limit. The levitation correction can be estimated~\cite{Haldane_97}
as
\begin{equation}
\delta E_N \sim \langle V^{\rm eff} \rangle_{\rm SP},
\label{delta_E_N} 
\end{equation}
where ``SP'' stands for saddle-points, i.e., the points where
\begin{equation}
(A)\:\: U_0^x = U_0^y = 0,\:\:
(B)\:\: U_0^{xx} U_0^{yy} - U_0^{xy} U_0^{yx} < 0
\label{AB}
\end{equation}
(the superscripts denote the partial derivatives). The rest of the
paper is devoted to the derivation of Eq.~(\ref{MF}) from
Eqs.~(\ref{V_eff}-\ref{delta_E_N}). I will assume that $U(x, y)$ is an
isotropic Gaussian random potential with zero mean.

It turns out that for each $k$, $Y_k$ and $Z_k$ are correlated with at most
one of the sets $\{U_0^x, U_0^y\}$ and $\{U_0^{xx}, U_0^{xy}, U_0^{yy}\}$.
Therefore, each time one needs to calculate either $\langle Y_k \rangle_A$ or
$\langle Y_k \rangle_B$, or simply the unrestricted average $\langle Y_k
\rangle$ (and similarly for $Z_k$).  The conditions ``$A$'' and ``$B$'' are
given by Eq.~(\ref{AB}). Notice that $Y_k$ and $Z_k$ are
bilinear in $U$.  This allows us to perform the $\langle\ldots\rangle_A$
averaging by means of the following general formula.  Let $X_1$ and $X_2$ be
linear in $U$, then
\begin{eqnarray}
\langle X_1 X_2 \rangle_A &=& \langle X_1 X_2 \rangle -
\displaystyle \langle U_0^x U_0^x\rangle^{-1}
\nonumber\\
&\times&
\left( \langle X_1 U_0^x\rangle \langle X_2 U_0^x\rangle +
\langle X_1 U_0^y\rangle \langle X_2 U_0^y\rangle \right).
\label{A_average} 
\end{eqnarray}
Similarly,
\begin{eqnarray}
\lefteqn{\langle X_1 X_2 \rangle_B = \langle X_1 X_2 \rangle
+ \langle U_0^{xx} U_0^{xx}\rangle^{-1}
(\langle X_1 U_0^{xy}\rangle \langle X_2 U_0^{xy}\rangle}
\nonumber\\
&-\frac12 \langle X_1 U_0^{xx}\rangle \langle X_2 U_0^{yy}\rangle -
\frac12 \langle X_1 U_0^{yy}\rangle \langle X_2 U_0^{xx}\rangle).&
\label{B_average} 
\end{eqnarray}
Equations~(\ref{A_average}) and (\ref{B_average}) can be obtained using
general properties of Gaussian potentials and isotropicity of $U$.
The $\langle\ldots\rangle_A$ averaging is needed for $Y_2$ and $Z_1$,
the $\langle\ldots\rangle_B$ averaging is needed for $Y_1$, $Y_3$, and $Z_2$.
All other $Y_k$'s and $Z_k$'s are to be averaged over the entire plane.
The calculation is trivial but lengthy, and so I will give only the final
result:
%
\[
\langle V^{\rm eff} \rangle_{\rm SP} = \frac{1}{m\omega_c^2}
\left[\frac{15 A_{1 3}^2 + A_{3 3}^2}{36 A_{0 4}} -
\frac{A_{2 2} (2 A_{0 2} + A_{2 2})}{4 A_{0 2}}\right],
\]
%
where $A_{kn}$ is defined as follows
\begin{equation}
A_{k n} = \int\! \frac{d^2 \bbox{q}}{(2 \pi)^2}\, \tilde{C}(q) J_0(q R)
J_k(q R)\, q^n,
\label{A_kn} 
\end{equation}
with $C$ being the correlator $C(r) =
\langle U(0) U(\bbox{r}) \rangle$. If $R_c \gg d$, then
\begin{equation}
\langle V^{\rm eff} \rangle_{\rm SP} \simeq
\frac{A_{0 2}}{4 m\omega_c^2} \sim \frac{W^2}{m \omega_c^2 R d}.
\label{V_SP II} 
\end{equation}
This concludes the derivation because Eqs.~(\ref{delta_E_N}) and
(\ref{V_SP II}) immediately give Eq.~(\ref{MF}).

The correction $(V^{\rm eff} - V_0)$ has quite peculiar properties: it
is typically positive at the points where the gradient squared
$(\bbox{\nabla}U_0)^2$ is smaller than its average value, typically
negative otherwise, and almost vanishes on average. Indeed, one can show
that the unrestricted spatial average of $V^{\rm eff}$ is $\langle
V^{\rm eff} \rangle = -A_{0 1}/(2 m\omega_c^2 R) \sim -\langle V^{\rm
eff} \rangle_{\rm SP} (d / R_c)^2$, which is much smaller than $\langle
V^{\rm eff} \rangle_{\rm SP}$ by the absolute value.

I speculate that the last property becomes important in the vicinity of
$B_c$ where the crossover from the drift to the diffusion occurs. In
the diffusive regime the trajectory of the guiding center is no longer
bound to the level line $V^{\rm eff} = {\rm const}$ but samples the
entire area. Thus $\langle V^{\rm eff} \rangle_{\rm SP}$ should approach
$\langle V^{\rm eff} \rangle$. The latter is indistinguishable from zero
within the accuracy of such an unrigorous argument. Thus, I expect the
ultimate downfall of the solid curve in Fig.~\ref{plot} near its
termination point.

Concluding this paper, let us emphasize that the discrepancy
between KL's formula~(\ref{KL}) and Eqs.~(\ref{Haldane})
and (\ref{MF}) does not contradict to Eq.~(\ref{scaling}), which
comes from the scaling arguments. Indeed, for $B < B_c$ the
``classical'' or the ``unrenormalized'' Hall conductance is determined not
by the average density $n$ but by the density $n_p$
{\em near the percolation contour\/} which is the area responsible
for the transport; therefore,
\begin{equation}
      \sigma_{xy} \simeq \frac{e^2}{m\omega_c} n_p.
\label{non_Drude}
\end{equation}
Equation~(\ref{delta_E_N}) now follows from Eqs.~(\ref{scaling}),
(\ref{non_Drude}), and $n_p = (m / \pi \hbar^2) (E_F - V^{\rm eff}_p)$
(``$p$'' again means ``percolation''). At $B > B_c$ the percolation
contour is of no importance so that Eq.~(\ref{Drude}) becomes valid, and
presumably so does Eq.~(\ref{KL}) as well.
 
In conclusion, I showed that the relative size $\delta E_N / E_N^\infty$ of
the weak levitation effect is always much smaller than one. Moreover, it
is expected to decrease near its termination point $B = B_c$. This strongly
suggests that this effect has nothing to do with the strong levitation
predicted by Khmelnitskii~\cite{Khmelnitskii} and
Laughlin.~\cite{Laughlin}

Finally, I should also mention that $\delta E_N$ can be measured
experimentally by charting the global phase diagram of the quantum Hall
effect,~\cite{Kivelson} the procedure initiated by Glozman {\it et
al.\/}~\cite{Glozman} Very interesting and puzzling findings have been
reported recently.~\cite{Kravchenko,Furneaux,Song} However, it seems that
the electron-electron interaction plays a crucial role in the observed
phenomena. This complicated issue is beyond the scope of the present
paper.

This work is supported by the DOE Grant No. DE-FG02-90ER40542, by
the University of Minnesota's Doctoral Dissertation Fellowship, and by
NSF Grant No. DMR-9616880. The idea for this work and some of the
intermediate formulas have come to life in the course of collaboration
with A.~Yu.~Dobin, V.~I.~Perel, and B. I. Shklovskii on
Ref.~\onlinecite{Fogler}. I thank them and also A.~I.~Larkin for useful
discussions. I especially thank B.~I.~Shklovskii for comments on the
manuscript.

\end{multicols}
\end{document}